\documentclass[sigconf]{acmart}

\usepackage{xcolor, graphicx, float}
\usepackage{colortbl}
\usepackage{listings}

\copyrightyear{2026}
\acmYear{2026}
\setcopyright{cc}
\setcctype{by}
\acmConference[ReCode '26]{1st Workshop on Code Translation, Transformation, and Modernization}{April 12--18, 2026}{Rio de Janeiro, Brazil}
\acmBooktitle{1st Workshop on Code Translation, Transformation, and Modernization (ReCode '26), April 12--18, 2026, Rio de Janeiro, Brazil}
\acmDOI{10.1145/3786180.3788313}
\acmISBN{979-8-4007-2411-4}

\title{CNnotator: LLM-Guided Memory Safety Annotation Synthesis}

\author{Twain Byrnes}
\authornote{Work completed while a Galois intern.}
\affiliation{%
  \institution{Carnegie Mellon University}
  \city{Pittsburgh}
  \state{PA}
  \country{USA}
}
\email{binarynewts@cmu.edu}

\author{Mike Dodds}
\affiliation{%
  \institution{Galois, Inc.}
  \city{Portland}
  \state{OR}
  \country{USA}
}
\email{miked@galois.com}



\begin{document}
\begin{abstract} 


Memory safety errors account for a large proportion of security bugs in systems written in C; modern languages such as Java and Rust prevent such bugs because they are memory-safe by design. To migrate systems to safer languages or identify memory errors, we must first determine how legacy code manipulates memory. This information is only represented implicitly in such code.

In many cases, memory usage patterns are merely tedious for humans to figure out, rather than truly difficult. In this work, we ask if large language models (LLMs) can perform this task by having them synthesize annotations representing memory usage as specifications in CN, a hybrid testing/verification tool. Our tool, CNnotator, uses LLMs to automatically generate and test CN specifications. We find that current models are able to generate CN specifications for small-to-medium C programs, with the OpenAI o3 reasoning model achieving a 90\% success rate on first attempts and 97\% overall success, while the chat model GPT-4o correctly annotates 65\% of first attempts. These results suggest AI-assisted annotation is becoming practical for real-world C codebases. 


\end{abstract} 
\maketitle

\section{Introduction} 

Formal methods and AI complement each other well: neural AIs are good at \emph{guessing} while formal methods tools are good at \emph{checking}. In fact, many formal methods problems can be phrased as guess-and-check loops. For example:

\begin{center}
\begin{tabular}{rcl}
\emph{Guess:} & & \emph{Check:} \\
\hline
program & $\rightarrow$ & specification \\
type signature & $\rightarrow$ & type checker \\
variable assignment & $\rightarrow$ & SAT solver \\
proof script & $\rightarrow$ & ITP tool \\
\end{tabular}
\end{center} 

Using a trustworthy formal checker, we don't need to trust the guesser. This approach neatly sidesteps the problems of AI unreliability and lack of interpretability. This paper describes \emph{CNnotator}, a tool which targets one such guess-and-check problem: figuring out how C programs use memory.\footnote{Portions of this paper previously appeared on the Galois blog \cite{cnnotator-blog-2025}.}

\paragraph{C and CN}

Memory safety issues cause a large proportion of security bugs. For example, Google estimated in 2020 that 70\% of Chromium security bugs arose from memory safety errors \cite{chromium-memory-safety}. In older languages such as C and C++, memory management is tedious to get right, easy to get subtly wrong, and can cause critical security errors that are nearly impossible to detect via unit testing. 

Modern languages, like Java and Rust, have built-in mechanisms that prevent nearly all memory safety bugs, with limited escape hatches for complex memory usage patterns. For users, these mechanisms are mostly automated, and many kinds of memory use can be handled by simple patterns. In C and C++, memory usage patterns cannot be explicitly stated or tested, and therefore they are easy to get wrong. But if we have ways of adding information that can cover simple usage patterns, perhaps we can verify a large amount of C and C++ code memory safe as well? 

Rather than annotate the code as is, we might transpile it to a memory-safe language. There are several approaches to this translation problem. Symbolic translation methods, like C2Rust \cite{c2rust}, Corrode \cite{corrode}, and Citrus \cite{citrus}, translate C to Rust, but produce unsafe, unidiomatic Rust with the expectation that guaranteed-safe, idiomatic code would require additional manual intervention. On the other hand, AI models can perform transpilation; the CRUST-bench benchmark shows that LLMs can generate idiomatic Rust code they claim to be valid and equivalent \cite{crustbench,crustbench-website}, but without formal guarantees. C2SaferRust combines symbolic and AI approaches \cite{c2saferrust}: idiomatic code from the LLM, correctness from the formal methods. All approaches must first understand memory patterns to verify correct translation.

This is one of the biggest challenges in modernizing or translating C code: figuring out how a given program manipulates memory. C programs do many complex things with memory, and this information is only implicit in the code itself. While tedious for humans to extract, these patterns are often simple enough for AI models to infer.

In this project, we built CNnotator, a tool which uses an LLM to determine what individual C functions do with memory. We represent memory usage information using CN \cite{cn-popl-2023}, a contract language similar to Frama-C \cite{framac}, but built on separation logic rather than standard Hoare logic. A contract in CN specifies a precondition and postcondition, and the CN tool then checks whether the contract holds. This testing can be performed by either an SMT solver \emph{or} by automated fuzzing/property-based testing over the contract. 

CN has two features that make it particularly useful for our purposes, compared to verification systems such as Frama-C:
\begin{enumerate}
  \item \emph{Memory representation.} CN specifications explicitly represent the structure of memory, meaning that specifications represent similar memory usage information to Rust's borrow checker (it inherits these ideas from \emph{separation logic} \cite{reynolds-seplogic-2002}). In particular, CN enforces memory safety through strict ownership requirements for pointers that are passed around, allocated, and freed. These properties are nearly identical to those that safe Rust enforces by default.
  \item \emph{Automated testing.} A CN specification can be exercised as a runnable test with no further human effort. The Fulminate backend \cite{fulminate-popl-2025} instruments the annotated C code, compiling the specification into runtime checks of internal assertions and postconditions and tracking memory footprints, so testing detects memory-safety violations as well as input-output errors. Building on Fulminate, the Bennet framework \cite{bennet-oopsla-2025} derives random input generators directly from the precondition, producing concrete heap states that satisfy it---including for heap-manipulating functions with non-trivial ownership preconditions, where hand-written generators would otherwise be required.
\end{enumerate}

For example, the CN annotations below specify that the function takes \emph{ownership} of the integer at \verb|p| (via \verb|RW<int>(p)|), increments it by one, and returns ownership upon completion. CN can automatically convert this specification into test cases that allocate memory, execute the function, and check that the postconditions hold.

\begin{lstlisting}[language=c]
void incr(int *p)
/*@ requires take x = RW<int>(p);
    ensures  take y = RW<int>(p);
             y == x + 1i32; @*/
{
  (*p)++;
}
\end{lstlisting}

There are broadly two kinds of CN annotations: those asserting functional correctness, and those asserting memory safety. In the above example, the first two requirements (those involving read-write permissions) are memory safety annotations. Without the memory safety annotations, CN cannot establish that the code executes without undefined behavior. The last condition asserts functional correctness, describing behavior during normal operation. CNnotator focuses only on memory safety annotations, not functional correctness, which is more subjective.

CN also offers SMT-backed verification, but we used only testing. Each test demonstrates memory safety for that specific run, giving concrete counterexamples on failure while avoiding internal annotations like loop invariants. 

\paragraph{CNnotator experiments}

We built our prototype tool, CNnotator, to test whether LLMs could generate CN specifications representing memory usage. At a high level, CNnotator follows the guess-and-check pattern: 

\begin{enumerate} 

  \item The AI model \emph{guesses} a possible CN specification for a given C function. 

  \item CN's testing backend \emph{checks} that the code passes one hundred automatically generated tests, with Bennet generating concrete heap states from the precondition and Fulminate checking the specification at runtime.

\end{enumerate} 

We experimented with several versions of CNnotator. Even the simplest version of this architecture was able to generate specifications for small-to-medium programs, and simple optimization techniques were effective at improving success rates. It seems plausible that, if AI capabilities continue to increase, we will be able to use similar techniques on larger and more complex programs in future. 







\paragraph{Paper structure}

The remainder of this paper is organized as follows. Section~\ref{sec:design} describes the design of CNnotator. Section~\ref{sec:benchmark} describes our benchmark suite. Section~\ref{sec:results} presents experimental results across five OpenAI models. Section~\ref{sec:conclusion} discusses tradeoffs and concludes. 
\begin{figure*}[t]
    \centering
    \includegraphics[width=0.84\textwidth]{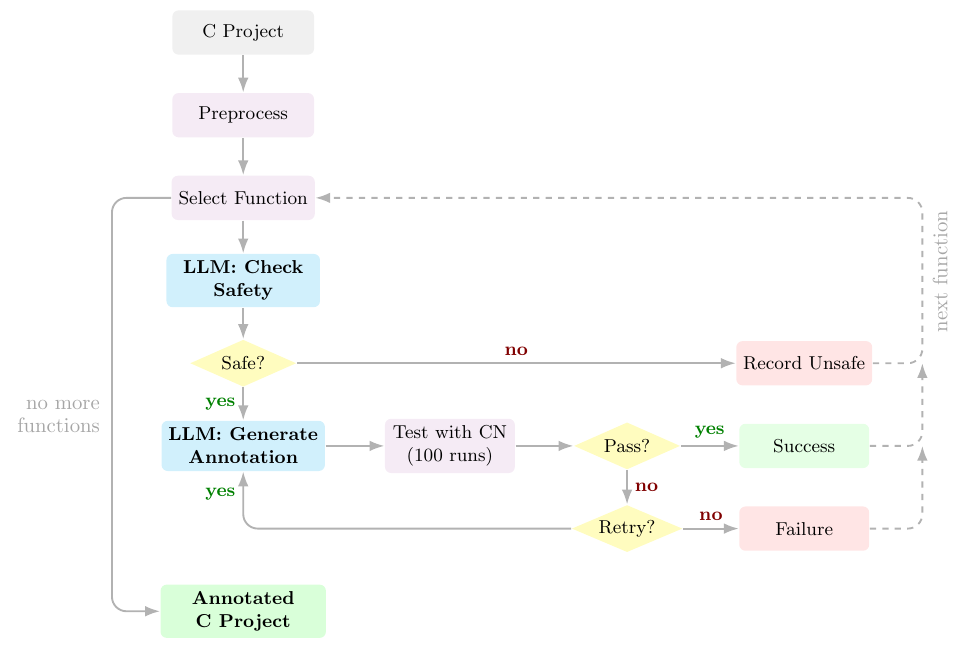}
    \caption{CNnotator workflow. The tool preprocesses C files (merging and adding CN macros), then iterates through each function. For each function, the LLM first checks for obvious memory safety bugs (e.g., use-after-free). If not obviously unsafe, it generates a CN annotation which is tested against 100 generated memory states. Failure tests trigger refinement with error feedback for up to six iterations.}
    \Description[CNnotator workflow diagram]{A flowchart showing the CNnotator workflow: The tool preprocesses C files (merging and adding CN macros), then iterates through each function. For each function, the LLM first checks for obvious memory safety bugs (e.g., use-after-free). If not obviously unsafe, it generates a CN annotation which is tested against 100 generated memory states. Failure tests trigger refinement with error feedback for up to six iterations.}
    \label{fig:workflow}
\end{figure*}
 
\section{CNnotator Tool Design}
\label{sec:design}

CNnotator is structured as a loop (Figure~\ref{fig:workflow} shows a simplified workflow): 

\begin{enumerate}
    \item \label{step:pick} Pick a C function dealing with memory ownership.
    \item \label{step:generate} Ask the LLM to generate a CN annotation for that function.
    \item \label{step:test} Test the annotation using the CN testing backend.
    \item \label{step:iterate} Depending on the result of testing, either refine the annotation, or go back to step~\ref{step:pick} and choose another function. 
\end{enumerate}

We implement CNnotator as a command line tool written in Python, with agent control flow and interactions handled by the LangChain and LangGraph libraries \cite{langchain, langgraph}. We use the Treesitter library \cite{treesitter} to parse the target C code and extract an abstract syntax tree (AST).

In preprocessing, we collate all C files in the project into one, so the LLM has full context for interdependent functions. We also add the necessary CN libraries and define macros so that allocation and deallocation work correctly with CN.

In step \ref{step:pick}, we iterate over the functions in the AST one at a time. We start from the bottom of the file to avoid invalidating AST positions.

In step \ref{step:generate}, we construct a prompt for the LLM based on the following template: 

\begin{center}
\begin{tabular}{|p{0.9\linewidth}|}
\multicolumn{1}{l}{\small \textbf{CNnotator prompt template:}} \\
\hline
\emph{CN annotation guide.} (400 lines approx)
Short guide for how to write CN annotations. Since LLMs have not been trained on CN syntax by default, we provide a tutorial to reference. The guide's structure and much of its text is scraped from the CN Tutorial \cite{cn-tutorial}, with small additions to cover common failure cases we observed. \\
\hline
\emph{Project code.} The source code for the whole C project. \\
\hline
\emph{Target function.} The specific function to annotate. \\
\hline
\emph{Formatting instructions.} A request to generate JSON output. \\
\hline
\end{tabular}
\end{center} 

In step \ref{step:test}, we inject the generated annotation into the target C code. We then test the annotation using CN's test synthesis capability. This runs the specification against many generated memory states. We test for 100 valid inputs, which is the default number of runs CN uses for its property-based testing feature.

In step \ref{step:iterate}, we process the result of running the tests, and choose one of two actions: 
\begin{enumerate} 

    \item If the test passes, we have completed specification generation for this function. We then return to the start of the loop and pick another function. 

    \item If it fails, we iterate on the generated annotation. We construct a new prompt consisting of the error, and for some common errors, a hint about how to rectify it. We then test the new annotation again up to $n$ times for any user-specified $n$. For our evaluation,
    we chose six, as all of the LLMs we tested seemed to hit diminishing returns around there. Past six, they would either cycle through ideas or progress towards unhelpful ends.

\end{enumerate} 

Prior work has shown that prompting an LLM multiple times on the same input leads to higher coverage, scaling with the number of samples \cite{large-language-monkeys}. Accordingly, if CNnotator fails to generate an annotation with correct syntax on its first try, it retries up to two more times using the original prompt rather than the error message. We retry with the original prompt because CN syntax error messages are often unhelpful to LLMs: they typically report only ``unexpected character'' at a location, or rely on visual formatting that models cannot interpret. Empirically, we found that fresh attempts frequently succeed where error-guided iteration does not.

At the end of the loop, CNnotator provides the user with a fully
annotated C project, or a comment recording the failure state. CNnotator can fail to generate annotations in three possible situations:

\begin{enumerate}
    \item No valid CN annotation exists for the C function, but the code is safe (some safe functions lack valid CN specifications due to the complexities of C memory access);
    \item A valid CN annotation exists, but CNnotator cannot find it;
    \item The C code is inherently unsafe (usually containing a bug), in which case no valid annotation can exist.
\end{enumerate}

CNnotator attempts to distinguish the last case from the first two. After picking a function (Step 1) but before asking the LLM to generate a CN annotation (Step 2), the LLM is asked whether the function is inherently unsafe. If it determines that a function \textit{is} inherently unsafe (e.g., contains a use-after-free or double-free pattern), it inserts a comment declaring that no annotation could ensure safety and records its reasoning in a separate file so the user can see how to modify the original code for safety.\footnote{CN is not designed to prove that a function is unsafe, although in future we could ask the LLM to generate such evidence, or reuse CN counterexample traces for this purpose.}

\section{Benchmark}
\label{sec:benchmark}

We created a test suite consisting of 31 annotatable C functions across 28 files and three unannotatable (i.e., inherently unsafe) C functions across three additional files. Of the 31 annotatable functions, 14 were drawn from the CN benchmark suite~\cite{cn-tutorial}. We included all of the functions in the benchmark suite that dealt with memory manipulation. These functions test basic CN features without complex memory patterns. 

We added 17 harder functions with subtle memory safety requirements, which were generated by Claude Sonnet and o3 as adversarial test cases. We removed comments from these generated functions and introduced additional complexity to challenge the annotation process. The three unannotatable functions contain deliberate memory-safety violations including double-free and use-after-free bugs, and were created manually to test CNnotator's ability to detect inherently unsafe code.

\paragraph{Benchmark Examples}
Below is a function from the benchmark set that loops through an array and sets each value equal to \texttt{7}:

\begin{lstlisting}[language=c]
void array_3(int *arr, int n) {
  int i = 0;
  while (i < n) {
    *(arr + i) = 7;
    i++;
  };
  return;
}
\end{lstlisting}

To annotate this function, CNnotator needs to recognize that the passed-in array \texttt{int *arr} is accessed for indices up to \texttt{n}, and thus must take ownership of array elements \texttt{0} through \texttt{n-1}.

In the next example, the array is accessed for values up to \texttt{ARRAY\_SIZE}. CN cannot use macros in annotations, so CNnotator must recognize that \texttt{ARRAY\_SIZE} is defined at the top of the file and must specify ownership of the first 5 array elements. For the second function, it must make the same decision to replace \texttt{ARRAY\_SIZE} with 5, but must additionally take ownership of the global variable \texttt{globalMultiplier}.

\begin{lstlisting}[language=c]
#define ARRAY_SIZE 5

int globalMult = 3;

void initGlobalArray(int *globalArray) {
    int i;
    for (i = 0; i < ARRAY_SIZE; i++) {
        *(globalArray + i) = i + 1;
    }
}

void multiplyGlobalArray(int *globalArray) {
    int i;
    for (i = 0; i < ARRAY_SIZE; i++) {
        *(globalArray + i) *= globalMult;
    }
}
\end{lstlisting}

\paragraph{Test set limitations}

We constructed our test set to exercise core CN contract idioms and standard C memory patterns, rather than to represent real-world code at scale. Most examples are under 50 lines, with simple control flow. We did not attempt to exercise the more complex usage patterns that CN supports, such as casts between types and complex pointer arithmetic. Several CN limitations also prevent testing on some categories of production code:

\begin{itemize}
    \item CN does not support some C types, such as union types.
    \item Many standard library functions are undefined for CN.
    \item CN uses custom \texttt{alloc} and \texttt{free} implementations, with \texttt{free} requiring the size of memory being freed. We worked around this using macros rather than allowing CNnotator to modify function bodies.
\end{itemize}

Despite these limitations, our benchmarks do exercise arrays, loops, global variables, macros, and multi-function files with memory interdependencies between functions.

\section{Results}
\label{sec:results}

\begin{figure*}[tb] 
\begin{tabular}{ll}
\includegraphics{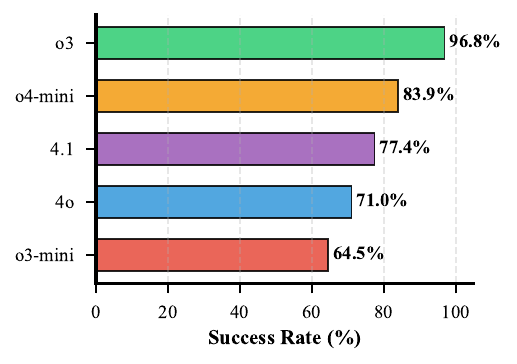}
&
\includegraphics{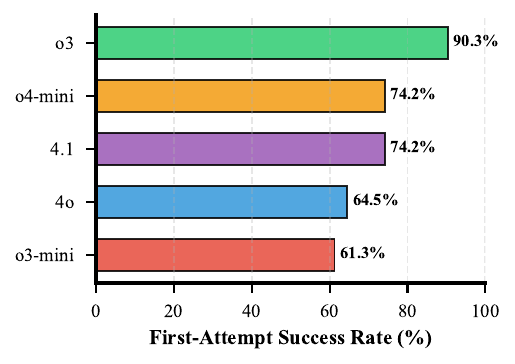}
\\
\includegraphics{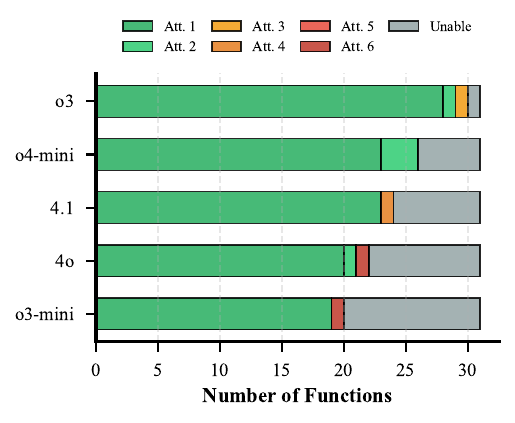}
&
\includegraphics{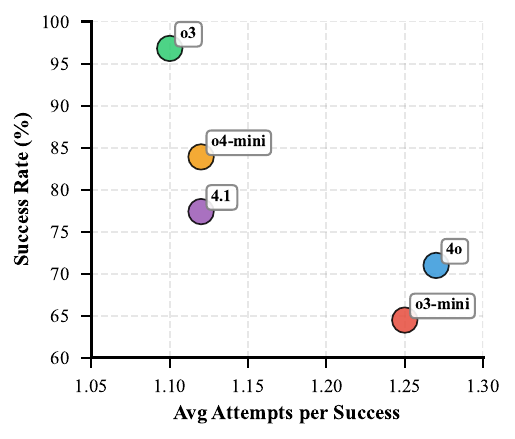}
\end{tabular} 

\caption{CNnotator performance across five LLM backends on 31 C functions. (a) Overall success rate. (b) First-attempt success rate without error correction. (c) Distribution of functions by attempts required (green: early success, red: multiple attempts or failure). (d) Efficiency frontier showing success rate versus average attempts. o3 achieves the best performance with 96.8\% overall success and 90.3\% first-attempt success.}
\Description[CNnotator Results]{CNnotator performance across five LLM backends on 31 C functions. (a) Overall success rate. (b) First-attempt success rate without error correction. (c) Distribution of functions by attempts required (green: early success, red: multiple attempts or failure). (d) Efficiency frontier showing success rate versus average attempts. o3 achieves the best performance with 96.8\% overall success and 90.3\% first-attempt success.}
\label{fig:cnnotator_results_vis}
\end{figure*}

Table~\ref{tab:cnnotator_results} presents CNnotator's performance across five LLM backends, while Figure~\ref{fig:cnnotator_results_vis} visualizes these results through multiple perspectives: overall success rates (a), first-attempt success without error correction (b), distribution of attempts required (c), and the efficiency frontier (d). 


\begin{table}[t]
\centering
\begin{tabular}{lrrr}
\toprule
 & Success \% & 1st Try \% & Avg. \# Attempts \\
\midrule
o3 & 96.8\% & 90.3\% & 1.10 \\
o4-mini & 83.9\% & 74.2\% & 1.12 \\
GPT-4.1 & 77.4\% & 74.2\% & 1.12 \\
GPT-4o & 71.0\% & 64.5\% & 1.27 \\
o3-mini & 64.5\% & 61.3\% & 1.25 \\
\bottomrule
\end{tabular}

\vspace{0.5em}

\small
\textit{Note: Success \% shows the percentage of functions successfully annotated. 1st Try \% shows functions that succeeded on the first attempt without errors. Avg Attempts is calculated only over successful functions.}

\vspace{0.7em}
\caption{CNnotator performance across LLM backends.}
\label{tab:cnnotator_results}
\end{table}

\paragraph{Testing setup}

For each benchmark file, we initialized a fresh LLM conversation to prevent information leaking between runs. However, within each file, the LLM maintained its session context, since proper annotations for functions with memory interdependencies may require knowledge of related annotations. CNnotator provides the complete project context to the LLM, enabling it to infer memory relationships and generate better specifications. Table~\ref{tab:cnnotator_results} reports only attempts to successfully annotate the annotatable functions; unannotatable functions are excluded from these metrics.

\paragraph{Performance}

In Table~\ref{tab:cnnotator_results}, o3, o3-mini, and o4-mini are reasoning models; GPT-4o and GPT-4.1 are chat models.

CNnotator with o3 achieves the best performance across all metrics in Table~\ref{tab:cnnotator_results}. It successfully annotates 30 of 31 functions (96.8\% overall success), with 28 succeeding on the first attempt (90.3\% first-attempt success rate). Only one function remains unannotated after the maximum number of attempts. In other words, CNnotator annotates 90\% of the test suite on the first attempt and reaches near-complete coverage with minimal iteration. CNnotator handles both simple ownership patterns and more complex functions with loops, arrays, global variables, and macros.

All tested GPT models successfully annotated the majority of functions on the first attempt, even chat models not specifically designed for complex reasoning tasks. This aligns with prior research demonstrating LLMs' strong comprehension of C code semantics~\cite{fang2024largelanguagemodelscode, ma2024lmsunderstandingcodesyntax}. Interestingly, GPT-4o (released August 2024, the oldest model tested) annotated 65\% of functions on first attempt, outperforming the newer o3-mini and matching o4-mini. This suggests model scale may partially compensate for lacking reasoning-specific optimizations.

As stated in Section~\ref{sec:design}, if CNnotator's first attempt at an annotation contains incorrect syntax, we retry the initial query up to three times instead of iterating on the broken annotation. We consider a proper annotation on any of these ``first attempts'' as only taking one attempt, as broken syntax does not yield useful hints. This exploits LLM nondeterminism rather than reasoning.

\paragraph{Failure Modes}

There are three failure modes for annotation. The first is inability to generate a passing annotation, which is automatically detected. The second failure mode is the generation of an annotation that passes the round of property-based testing, but the generated specification is still incorrect for some inputs. For our best-performing model (o3), all annotated functions were manually inspected for correctness. For other models, we inspected the adversarial benchmarks, which were more challenging. Manual inspection showed that all annotations deemed correct by CN's property-based testing feature were indeed correct on all inputs.

The third failure mode is that the LLM may ``cheat'' by generating an over-permissive annotation. For example, if a memory-safe function is passed a pointer to an array and the length of that array, a CN annotation requiring ownership of twice the length would pass test cases, though it requires more than is necessary and does not match actual usage of the function. Since this kind of pattern depends on the surrounding code, we feed the whole program to CNnotator so that it may make a more informed decision. These usage patterns were manually investigated along with the second failure mode described above. Though different requirements were found across different runs, none violated usage of the function within the file passed in.

Our benchmark includes several inherently unsafe functions containing use-after-free and double-free bugs. All models detected these immediately and declined to annotate them. However, subtler bugs (such as out-of-bounds array access) would be harder to detect. As noted above (failure mode three), CNnotator might generate an over-approximating annotation that requires ownership of more memory than strictly necessary. We consider such over-approximation acceptable: it still yields a valid specification, and the ``correct'' memory footprint often depends on calling context.

\section{Discussion}
\label{sec:conclusion}

\paragraph{Framework Tradeoffs}
As an experiment, we also built \emph{CNnotatorCC}, reimplementing CNnotator using the agentic framework Claude Code~\cite{claude-code}. Because Claude Code edits files directly and manages its own control flow, we built CNnotatorCC in less than a day, compared to several weeks for the original tool. In limited manual testing, CNnotatorCC with Claude Sonnet performed comparably to CNnotator with o3.

However, agentic frameworks require more trust than structured tools. CNnotator inserts annotations via a deterministic Python script; Claude Code may edit code in unexpected ways, including modifying the function under test rather than admitting it cannot find an annotation. We also found Claude Code unreliable at self-reporting: in one case, it reported 3 attempts when logs showed 8, later claiming it had discounted attempts it ``deemed unimportant.'' For unattended automation, structured architectures currently offer stronger guarantees.

\paragraph{Future Work}
CNnotator generates memory safety specifications in CN, a separation-logic-based language. This could facilitate direct translation to safe Rust, skipping intermediate unsafe Rust. In future work, we plan to translate C code with CN annotations directly to safe Rust, and evaluate when and how these annotations are helpful. Another potential direction is experimenting with more LLM providers. We expect annotation capabilities to improve over time, as LLM reasoning improves.

\paragraph{Conclusion}
CNnotator can help determine whether legacy C code is memory safe, using CN's property-based testing to generate counterexamples that guide annotation refinement. It can synthesize annotations for multi-function, multi-file C projects with various data structures.

Combining LLMs with formal methods enables new approaches to code modernization. As shown by our experiments, LLM outputs are not to be trusted. However, formal checkers can validate them without requiring trust. CNnotator contributes to C-to-Rust translation research, such as DARPA's TRACTOR program~\cite{darpa-tractor}, by automating memory safety annotation, and CNnotator can also help modernize C code without full translation.

\bibliographystyle{ACM-Reference-Format}
\bibliography{references}

\end{document}